\pdfoutput=1
\documentclass[11pt,a4paper]{article}
\usepackage{amsmath,amssymb}
\usepackage{braket}
\usepackage{url}
\usepackage{mathtools}
\usepackage{amsmath}              
\usepackage{graphics,graphicx,epsfig,ulem} 
\usepackage{subfigure,subfig}
\usepackage{feynmp}
\usepackage{slashed}
\usepackage{color}

\DeclareMathAlphabet{\mathbold}{U}{zeur}{b}{n}

\renewcommand\[{\left[}
\renewcommand\]{\right]}

\def\beq{\begin{equation}}
\def\eeq{\end{equation}}
\def\[{\begin{equation}}
\def\]{\end{equation}}

\begin{document}
\numberwithin{equation}{section}

\title{
{\normalsize  \mbox{}\hfill IPPP/17/36}\\
\vspace{2.5cm}
\Large{\textbf{Multiparticle production in the large $\lambda n$ limit:\\ 
Realising Higgsplosion in a scalar QFT}}}

\author{Valentin V. Khoze 
 \\[4ex]
  \small{\it Institute for Particle Physics Phenomenology, Department of Physics} \\
  \small{\it Durham University, Durham DH1 3LE, United Kingdom}\\
  [1ex]
  \small{\tt valya.khoze@durham.ac.uk }\\
  [0.8ex]
}

\date{}
\maketitle

\begin{abstract}
\noindent In a scalar theory which we use as a simplified model for the Higgs sector, we
adopt the semiclassical formalism of Son for computations of $n$-particle 
production cross-sections in the high-multiplicity $n\to \infty$ weak-coupling $\lambda \to 0$ regime
with the value of $\lambda n$ held fixed and large.
The approach relies on the use of singular classical solutions to a certain boundary value problem. 
In the past this formalism has been successfully used and verified in computations of perturbative multi-particle processes 
at tree-level, and also at the next-to-leading order level in the small $\lambda n$ expansion near the multi-particle
mass threshold. We apply this singular solutions formalism in the regime of ultra-high multiplicities where $\lambda n \gg 1$,
and compute the leading positive $\sim  n\,\sqrt{\lambda n}$ contribution to the exponent of the multi-particle rate
in this large $\lambda n$ limit. The computation is carried out near the multi-particle mass threshold 
where the multiplicity $n$ approaches its maximal value allowed by kinematics. This calculation relies on the 
idea of Gorsky and Voloshin to use a thin wall approximation for the singular solutions that resemble critical bubbles.
This approximation is justified in precisely the high-multiplicity $\sqrt{\lambda n} \to \infty$ regime of interest.
Based on our results we show that the scalar theory with a spontaneous symmetry breaking used here
as a simplified model for the Higgs sector, is very likely to realise the high-energy Higgsplosion phenomenon.
\end{abstract}

\bigskip
\thispagestyle{empty}
\setcounter{page}{0}

\newpage


\section{Introduction}
\label{sec:intro}
\medskip

The discovery of a light Higgs boson at the Large Hadron Collider \cite{Aad:2012tfa,Chatrchyan:2012xdj},
taken together with the apparent lack of any evidence for additional 
beyond the Standard Model degrees of freedom at energies accessible by current experiments,
leaves us with a fundamental problem of how to stabilise the Higgs mass.  With a safe assumption that the Standard Model
does not account for all microscopic interactions in nature and that the more complete theory is likely to
include some super-heavy degrees of freedom\footnote{These could be 
for example heavy vectors and scalars of a Grand Unified Theory, 
heavy sterile neutrinos responsible for a thermal vanilla leptogenesis, flavons, or states with the Planck or string-scale masses.}
we are led to the well-known Hierarchy or the fine-tuning problem for the Higgs mass.  Quantum corrections to the Higgs mass
induced by the scale of new physics in the UV push the Standard Model Higgs boson mass parameter into the UV domain,
unless there is an underlying symmetry reason that the quantum effects cancel among each other, or are not present to start with.

\bigskip

One very recent proposal for addressing the Hierarchy problem that does not rely on supersymmetry or Higgs compositeness,
is the Higgsplosion mechanism introduced in Ref.~\cite{Higgsplosion}. The main idea of the approach is to destroy all the
super-heavy states $X$ by allowing them to rapidly decay into multiple Higgs bosons $X \to n\times h$ at energy scales
much below their mass $M_X$. In other words, one aims to have the multi-particle decay widths 
$\Gamma_{X\to n\times h}$ to exceed $M_X$
at energies $\sqrt{s_\star} \ll M_X$. In this sense, the heavy $X$ states become unrealised as particle states, they decay
faster than they form, and in practical calculations, the loop integrals involving loops of virtual $X$ fields are effectively cut-off 
at the relatively low scale $\sqrt{s_\star} \ll M_X$.

\bigskip

The aim of the present article is to show that the Higgsplosion mechanism can be realised in 
simple quantum field theoretical settings. As in Ref.~\cite{Higgsplosion} we will concentrate on a model
with a single real scalar degree of freedom $h(x)$,
\[{\cal L} \,= \, \frac{1}{2}\, \partial^\mu h \, \partial_\mu h\, -\,  \frac{\lambda}{4} \left( h^2 - v^2\right)^2
\,. \label{eq:L}
\]
This theory is a reduction of the SM Higgs sector in the unitary gauge to a single scalar field
$h(x)$ which for our purposes we take to be stable, so there are no decays into fermions, and we have also decoupled 
all vector bosons, etc. 
The vacuum expectation value $v$ results in a spontaneous symmetry breaking of the $h\to -h$
discrete symmetry and the field  $\varphi(x) = h(x)-v$ 
describes the boson of mass $M_h = \sqrt{2\lambda}\,v$.
From now on we will treat \eqref{eq:L}
as the simplified model description of the self-interacting Higgs sector and will ignore effects of
other interactions of the Higgs with the Standard Model vectors and fermions. Clearly the effects of such interactions will 
ultimately need to be understood and estimated  for a more realistic phenomenological treatment. 
Here we will stick with a simpler goal -- which is to demonstrate  that 
the  concept of Higgsplosion can be realised in a concrete simple scalar field theory example. In the Discussion section we will
briefly comment on the more general cases.

\bigskip

The aim of this paper is to compute the multi-boson production rate in the large $\lambda n$ limit,
where $\lambda$ is the coupling constant and $n$ is the particle number in the final state.
On the technical side, the idea which makes this calculation possible, is to 
combine the semiclassical formalism developed by Son in Ref.~\cite{Son} 
based on singular classical solutions  with the 
approach of Gorsky and Voloshin \cite{GV} which will allow us to search for these solutions
in the form of thin walled singular bubbles.

\bigskip

This paper is organised as follows. In section \ref{sec:2}, 
following a brief recollection of known results for multi-particle amplitudes, we will summarise the semiclassical approach of
\cite{Son} aimed at computing the cross-sections for such processes at very
high energies. In section \ref{sec:3} we will continue with this semiclassical technique and will relate it to 
the problem of finding extrema of Euclidean actions computed on singular surfaces. 
This problem will be addressed and solved in section \ref{sec:4}  using the thin-wall approximation 
in the large final state multiplicity limit. 
There we will employ the approach of \cite{GV} developed for thin-wall bubbles.
Finally, in section \ref{sec:5} we will provide a detailed discussion of the main results, 
their consequence for the Higgsplosion picture \cite{Higgsplosion} and comments on future directions.

\medskip
\section{Semiclassical approach for multi-particle production}
\label{sec:2}

In the scattering processes at very high energies, production of large numbers of particles in the final state 
becomes possible. We will concentrate on such processes in a scalar field theory. These processes
were studied in some detail in the  literature and we refer the reader to a selection of papers
\cite{Cornwall:1990hh,Goldberg:1990qk,Voloshin:1992mz,Brown,Argyres:1992np,Voloshin:1992rr,Voloshin:1992nu,Smith:1992rq,LRST,GV,Son,LRT,VVK2,Jaeckel:2014lya,VVK3}
and references therein.

\bigskip

To start, we consider the leading order tree-level $n$-point scattering amplitude 
computed on the $n$-particle mass thresholds. This is the kinematics regime where all
final state particles are produced at rest. These amplitudes for all $n$ are conveniently assembled 
into a single object -- the amplitude generating function -- which at tree-level is described by 
a particular solution of the Euler-Lagrange equations.
The classical solution which provides the generating function of tree-level amplitudes on multi-particle
mass thresholds in the model \eqref{eq:L} is given by~\cite{Brown},
\[
h_{\rm cl} (z_0;t) \,=\, v\, \left(\frac{1+z_0\, e^{iM_h t}/(2v)}{1-z_0\, e^{iM_h t}/(2v)}\right)\,,
\label{clas_sol}
\]
where $z_0$ is an auxiliary variable. It is easy to check with the direct substitution that the
expression in \eqref{clas_sol} does indeed satisfy the Euler-Lagrange equation resulting from our theory
Lagrangian \eqref{eq:L} for any value  of the $z_0$ parameter. It then follows 
that all $1\to n$ tree-level scattering amplitudes on the $n$-particle mass thresholds are given by the differentiation 
of $h_{\rm cl} (z_0;t)$ with respect to $z_0$,
\[
{\cal A}_{1\to n}\,=\, \langle n|S \phi(0)| 0\rangle \,=\,\, \left.\left(\frac{\partial}{\partial z_0}\right)^n h_{\rm cl} \,\right|_{z_0=0}
 \label{eq:A5}
\]
The classical solution in \eqref{clas_sol} can be thought of as a holomorphic function of the complex variable 
$z(t) = z_0\, e^{iM_h t}$,
\[
h_{\rm cl} (z(t)) \,=\,  v\,+\, 2v\,\sum_{n=1}^{\infty} \left(\frac{z(t)}{2v}\right)^n
\, ,
\label{gen-funh2}
\]
so that the amplitudes in \eqref{eq:A5} are given by the coefficients of the Taylor expansion in \eqref{gen-funh2} 
times $n!$ from differentiating $n$ times over $z$,
\[
{\cal A}_{1\to n}\,=\, 
\left.\left(\frac{\partial}{\partial z}\right)^n h_{\rm cl} \,\right|_{z=0}
\,=\, n!\, \left(\frac{1}{2v}\right)^{{n-1}}
\,,
\label{eq:ampln2}
\]
These formulae and the characteristic factorial growth of $n$-particle amplitudes,  ${\cal A}_{n} \sim \lambda^{n/2} n!$, 
form the essence of the elegant formalism pioneered by Brown in Ref.~\cite{Brown} that is 
based on solving classical equations of motion and bypasses the summation over individual Feynman diagrams. 
For more detail and derivations we refer the reader to the original paper \cite{Brown} or a review in Section 2 of Ref.~\cite{VVK1}. 

We now perform the Wick rotation from the real Minkowski time $t$ to the Euclidean time $t_{\rm Eucl} = i t$.
To use the same notation for the imaginary time variable as in \cite{Son} we will use the variable $\tau$ defined as,
\[ \tau\,:=\, -\, t_{\rm Eucl} \,=\, -\, i t\,.
\label{tau_def}
\]
Expressed as the function of the Wick-rotated time variable $\tau$, the classical solution \eqref{clas_sol} reads,
\[
h_{\rm cl} (\tau) \,=\, v\, \left(\frac{ 1\,+\,e^{-M_h (\tau-\tau_\infty)}}
{1\,-\, e^{-M_h (\tau-\tau_\infty)}}\right)\,,
\label{clas_sol2}
\]
where the $\tau_\infty$ parameter,  $\tau_\infty = \frac{1}{M_h} \log \left(\frac{z_0}{2v}\right)$, gives the location 
of the solution in time.
The sign convention in \eqref{tau_def} where $\tau$ is identified with the negative of the Euclidean time,
 implies that the early  time $t \to -\infty$ corresponding to the 
initial time, i.e. the incoming states, maps to $\tau \to +\infty$. 
In this limit the classical solution approaches the vacuum $h_{\rm cl} \to v$
with exponential accuracy, i.e. the corrections are ${\cal O}(e^{-M_h \tau})$.

The expression on the right hand side of \eqref{clas_sol2} has an obvious interpretation 
in terms of a singular domain wall located at $\tau=\tau_\infty$ that separates two domains of the field $h(\tau, \vec{x})$. 
The domain on the right of the wall 
$\tau \gg \tau_\infty$ has $h \sim +v$,
and the domain on the left of the wall, $\tau \ll \tau_\infty $, is characterised by $h \sim -v$.
The field configuration is singular at the position of the wall, $\tau=\tau_\infty$, for all values of $\vec{x}$, i.e. the singularity surface is 
flat (or uniform in space). The thickness of the wall is set by  $1/M_h$.

\bigskip

In Ref.~\cite{Son} Son proposed a semiclassical approach for computing multi-particle
cross-sections in a scalar QFT. This approach is quite general, as it works not only
for the leading-order tree-level processes, but is also capable of computing the higher-order quantum loop-level effects.
Furthermore the method is designed to provide probabilistic quantities, i.e. the rates or cross-sections, hence it 
goes beyond just the calculation of the amplitudes near or on the mass-thresholds by also taking  account of the integrations over the 
$n$-particle Lorentz-invariant phase space. 
The approach of \cite{Son} generalised to field theory the Landau WKB method \cite{Landau} 
for computing
matrix elements of certain generic local operators between the initial and final states with different 
energy eigenvalues. In our case, the initial state is a vacuum and the final state is the $n$-particle final state with $n\gg 1$.
It is known that to the leading exponential accuracy the transition rates computed using the Landau WKB method do not
depend on the specific form of the operator ${\cal O}(x)$ used to deform the initial state, if this deformation is not exponential. 
It is then similarly expected that 
the choice of the operator does not affect the transition rates in the QFT settings either, and the approach of  
\cite{Son} generalises the Landau WKB method to a scalar QFT using the path integral formalism.

The central quantity is the dimensionless probability rate ${\cal R}_n(E)$ for a local operator ${\cal O}$ 
to create $n$ particles of total energy $E$ from the vacuum. It is given by \cite{Son},
\[
{\cal R}_n(E)\,=\, \int d\Phi_n\, \langle 0| \,{\cal O}^\dagger \, S^\dagger\, P_E |n\rangle 
\langle n|\, P_E\, S\, {\cal O}\, |0\rangle\,,
\label{eq:RnE1}
\]
where the matrix element involves the operator ${\cal O}$  between the vacuum state $ |0\rangle$ and the $n$-particle state
of fixed energy $\langle n|\, P_E $ (here $P_E$ is the projection operator on states with fixed energy $E$), along with
the $S$ matrix to evolve between the initial and finial times. The matrix element is squared and integrated over the 
$n$-particle Lorentz-invariant phase space $\Phi_n$
\[ 
\int d\Phi_{n} \,=\, \frac{1}{n!}\,(2\pi)^4 \delta^{(4)}(P_{\rm in}-\sum_{j=1}^n p_j) \,
 \prod_{j=1}^n \int \frac{d^3 p_j}{(2\pi)^3\, 2 p_j^0} \,.
 \label{eq:phase-sp}
\]
Note that in our conventions 
the bosonic phase-space volume element \eqref{eq:phase-sp} 
includes the $1/n!$ symmetry factor for the production of the $n$ equivalent Higgs bosons.\footnote{Hence the $n$-particle 
cross-sections ${\cal R}_n(E)$ still retains a single factor of $n!$. Indeed, according to \eqref{eq:ampln2},
the amplitude squared contributes the factor of $(n!)^2$, and combining with the symmetry factor from the 
bosonic $n$-particle phase space 
we have ${\cal R}_n(E) \sim\, \frac{1}{n!} \, n! \, n! \sim \, n!$.
}

The local operator ${\cal O}$ appearing in the matrix elements in \eqref{eq:RnE1} is usually chosen to be \cite{Son}
\[ {\cal O} \,=\,{e^{j\,(h(0)-v)}}:= e^{j \phi(0)} \,,\]
where $j$ is a constant, and the limit $j\to 0$ is taken in the computation of the probability rates,
\[
{\cal R}_n(E)\,=\, \lim_{j\to 0} \int d\Phi_n\, \langle 0| \,{e^{j\phi(0)}}^\dagger \, S^\dagger\, P_E |n\rangle 
\langle n|\, P_E\, S\, {e^{j\phi(0)}}\, |0\rangle\,.
\label{eq:RnE}
\]
The cross-sections for few to many particles, $\sigma_{{\rm few}\to n} (E)$ as well as multi-particle partial decay 
rates $\Gamma_n(E)$ of a single particle state $X \to n\times h$, are determined by the exponential factor for
${\cal R}_n(E)$ in \eqref{eq:RnE} times a non-exponential pre-factor of appropriate dimensionality which is 
of no interest in a semiclassical approximation.

In the construction of~\cite{Son} the expression on the right hand side of  \eqref{eq:RnE} 
is represented as a functional integral, which is subsequently computed in the steepest descent approximation
for all integration variables. This is achieved and justified in the 
double-scaling weak-coupling / large-$n$ semiclassical limit:
\[ \lambda \to 0\,, \quad n\to \infty\,, \quad {\rm with}\quad
\lambda n = {\rm fixed}\,, \quad \varepsilon ={\rm fixed} \,.
\label{eq:limit}
\]
Here $\varepsilon$ denotes the average kinetic energy per particle per mass in the final state,
\[ \varepsilon \,=\, (E-nM_h)/(nM_h)\,.
\]
Holding $\varepsilon$ fixed implies that in the large-$n$ limit we are raising the total energy linearly with $n$.

\bigskip

The semiclassical result for the rate has the characteristic exponential form \cite{Son},
\[
{\cal R}_n(E)\,\simeq \, \exp \left[ W(E,n)\right],
\label{eq:ReW}
\]
where 
\[ W(E,n) \,=\, \frac{1}{\lambda}\, {\cal F}(\lambda n, \varepsilon)\,=\,
ET \,-\, n\theta \,-\, 2{\rm Im} S[h]\,.
\label{eq:Wdef}
\]
Let us now examine the structure of this result.
The function ${\cal F}(\lambda n, \varepsilon)$ appearing in \eqref{eq:Wdef}, is a function of two finite-valued arguments
while all the integrations in the path integral representation of ${\cal R}_n(E)$ in \eqref{eq:RnE}  were carried out 
and saturated by their saddle-point values in the large-$n$, large-$1/\lambda$ limit \eqref{eq:limit}.
At negative values of ${\cal F}(\lambda n, \varepsilon)$ the multi-particle rate ${\cal R}_n(E)$ is exponentially suppressed,
while if ${\cal F}(\lambda n, \varepsilon)$ crosses zero and becomes positive above some critical energy or multiplicity, the
multi-particle processes enter the Higgsplosion phase \cite{Higgsplosion}.

\bigskip

\noindent We now consider the terms appearing in the final expression in \eqref{eq:Wdef}.
First, the combination $-\, 2{\rm Im} S[h]$ follows from the $e^{-iS[h]^*}e^{iS[h]}$ factor 
in the product of the matrix elements in  \eqref{eq:RnE}. The integration contours and the 
resulting saddle-points in the steepest descent integration are complex-valued, 
hence $iS[h]-iS[h]^*\, =\, -\, 2{\rm Im} S[h]$
or equivalently $-2S_{\rm Eucl}[h]$ using the Euclidean notation.
Finally, the parameters $T$ and $\theta$ appearing on the right hand side of \eqref{eq:Wdef}  are the consequence 
of introducing projections onto the final states with defined values of the energy $E$ and the particle number $n$ 
in \eqref{eq:RnE}.

The function $W(E,n)$ in \eqref{eq:ReW}-\eqref{eq:Wdef} is computed on the saddle-point value of the path integral.
Prior to taking the $j\to 0$ limit in \eqref{eq:RnE}, the
saddle-point field configuration $h(x)$ is given by a particular solution to the classical equation
of motion with the singular 
source term $j(x)=\, j \delta^{(4)}(x)$ on the right hand side,
\[
\frac{\delta S}{\delta h(x)}\,=\, i\, j\, \delta^{(4)}(x)\,,
\label{sing_eqn}
\]
where $S=\int d^4 x {\cal L}$ is the action of the theory and $j$ is a constant.
After taking the limit $j\to 0$, the right hand side of the defining equation \eqref{sing_eqn} vanishes
but the required solution nevertheless remains singular at $x=0$ in Minkowski space.
The saddle-point solution also depends on the parameters $T$ and $\theta$, as will be explained below, while the 
overall expression $W(E,n)$ is independent of  $T$ and $\theta$. Hence,
\[ 
2\, \frac{\partial\,{\rm Im}S}{\partial T}\,=\, E\,, \qquad
2\, \frac{\partial\,{\rm Im}S}{\partial \theta}\,=\, -\,n\,,
\label{eqs:Tthetadef}
\]
and $W(E,n)$ is the Legendre transformation of the action $2{\rm Im}S$ 
with respect to $T$ and $\theta$.\footnote{Indeed, it follows from the definition of $W$ that
$\frac{\partial W}{\partial E}\,=\, T$ and 
$\frac{\partial W}{\partial n}\,=\, -\,\theta$.
The action $S[h]$ depends on the parameters $T$ and $\theta$ through the classical solution $h(x)$,
but in the final expression for $W(E,n)$ these parameters are traded for $E$ and $n$. }

Next step is to specify the boundary conditions of the solution $h(x)$ at  $t_{\rm in}\to -\infty$ and $t_{\rm fin}\to +\infty$.
At the initial and final time boundaries $h(x)$ satisfies the free Klein-Gordon equation, thus
\begin{eqnarray}
h(\vec{x},t)|_{t\to - \infty} &\to& v\,+\, 
\int \frac{d^3k}{(2\pi)^{3/2}} \frac{1}{\sqrt{2\omega_{\bf k}}}\,\, a^\dagger_{\bf k}\, e^{ik_\mu x^\mu} \,
\label{eq:limin}
\\
h(\vec{x},t)|_{t\to + \infty} &\to& v\,+\, 
\int \frac{d^3k}{(2\pi)^{3/2}} \frac{1}{\sqrt{2\omega_{\bf k}}}\left(c_{\bf k}\, e^{-ik_\mu x^\mu}\,+\, b^\dagger_{\bf k}\, e^{ik_\mu x^\mu}\right)\,.
\label{eq:limf}
\end{eqnarray}
where we used the standard notation $k_0=\omega_{\bf k}=\sqrt{M_h^2+{\bf k}^2}$
so that $e^{\pm i k_\mu x^\mu} = e^{\pm i(\omega_{\bf k} t \,-\, {\bf k} {\bf x})}$. \\
The $t \to -\infty$ boundary condition in Eq.~\eqref{eq:limin} 
contains only the positive frequency components
$\frac{1}{\sqrt{2\omega_{\bf k}}}\, a^\dagger_{\bf k} \, e^{-i\omega_{\bf k} |t|}$ and no negative frequency ones 
$\frac{1}{\sqrt{2\omega_{\bf k}}}\, a_{\bf k} \, e^{+i\omega_{\bf k} |t|}$.
In the second quantisation operator formalism, this condition
implements the requirement that there are no particles in the initial state, 
$ \langle 0|\, \int \frac{1}{\sqrt{2\omega_{\bf k}}}\, a^\dagger_{\bf k} \, e^{-i\omega_{\bf k} |t| } \, = \, 0$
since the creation operator $a^\dagger$ annihilates the bra-state vacuum $\langle 0|.$
The second boundary condition \eqref{eq:limf} at the final time $t\to +\infty$
contains both positive and negative frequency components.
Following \cite{Son} we parameterise its $c_{\bf k}$ coefficient  
in terms of the complex conjugate of its $b_{\bf k}^\dagger$ coefficient,
\[
c_{\bf k}\,=\, b_{\bf k}\,e^{\omega_{\bf k}T-\theta}\,.
\label{eq:cb}
\]
The solution is complex-valued since $c_{\bf k}\neq\, b_{\bf k}$,
and the corresponding parameters $T$ and $\theta$ are precisely those appearing in \eqref{eqs:Tthetadef}.

\bigskip

\noindent In summary, the equations~\eqref{sing_eqn}-\eqref{eq:cb} specify the boundary value problem for finding the
saddle-point configuration $\{h(x), T, \theta\}$ needed to compute the semiclassical rate ${\cal R}_n(E)$:


\begin{enumerate}
\item Solve the classical equation without the source-term,
\[
\frac{\delta S}{\delta h(x)}\,=\,0\,,
\nonumber
\]
by finding a complex-valued solution $h(x)$ with a point-like singularity at the origin
$x^\mu=0$ and regular everywhere else in Minkowski space.

\item Impose the initial and final-time boundary conditions,
\begin{eqnarray}
\lim_{t\to - \infty}\,h(x)  &=& v\,+\, 
\int \frac{d^3k}{(2\pi)^{3/2}} \frac{1}{\sqrt{2\omega_{\bf k}}}\,\, a^\dagger_{\bf k}\, e^{ik_\mu x^\mu} \,
\nonumber
\\
\lim_{t\to + \infty}\,h(x) &=& v\,+\, 
\int \frac{d^3k}{(2\pi)^{3/2}} \frac{1}{\sqrt{2\omega_{\bf k}}}\left( b_{\bf k}\,e^{\omega_{\bf k}T-\theta}
\, e^{-ik_\mu x^\mu}\,+\, b^\dagger_{\bf k}\, e^{ik_\mu x^\mu}\right)\,.
\nonumber
\end{eqnarray}

\item Compute the energy and the particle number using the 
$t\to +\infty$ asymptotics of $h(x)$,
\[
E \,=\, \int d^3 k \,\, \omega_{\bf k}\, b_{\bf k}^\dagger \, b^{}_{\bf k}\, e^{\omega_{\bf k}T-\theta}
\,, \qquad
n \,=\, \int d^3 k \,\, b_{\bf k}^\dagger \, b^{}_{\bf k}\, e^{\omega_{\bf k}T-\theta}\,.
\nonumber  
\]
At $t\to -\infty$ the energy and the particle number  are vanishing. 
The energy is conserved by regular solutions and changes discontinuously from $0$ to $E$ 
at the singularity at $t=0$.

 \item Eliminate the $T$ and $\theta$ parameters in favour of $E$ and $n$ using the expressions above.
 Finally, compute the function $W(E,n)$
  \[ W(E,n)  \,=\,
ET \,-\, n\theta \,-\, 2{\rm Im} S[h]
\nonumber 
\]
on the set $\{h(x), T, \theta\}$, and thus determine the semiclassical rate ${\cal R}_n(E) \,=\, \exp \left[ W(E,n)\right]$.
\end{enumerate}

\bigskip
\section{Solving the boundary-value problem by extremizing the \\
Euclidean action over the singular surfaces}
\label{sec:3}
\bigskip

One way to visualise the construction of the solution to the boundary-value problem outlined above,
is by starting with the specified expressions for $h(x)$ at the $t\to \pm \infty$ boundaries and classically
evolving them by numerically solving the equation of motion into the region of
finite $t$. We thus have two trial functions, one at $t<0$ and the 
second at $t>0$  which we would like to match at $t=0$.
The field configuration at $t<0$ is given by a regular classical solution $h_1(t,\vec{x})$ which 
satisfies the initial time boundary condition with the Fourier coefficient functions $a_{\bf k}$.
The second trial function, $h_2(t,\vec{x})$, is a regular classical solution on the Minkowski half-plane $t>0$
which is evolved from the final-time boundary condition with the coefficient functions $b_{\bf k}$.
One then contemplates scanning over the space of the functions $a_{\bf k}$ and $b_{\bf k}$ to achieve
the matching at $t=0$  between the two branches $h_1$ and $h_2$ of the solution, $h_1(\vec{x})=h_2(\vec{x})$, 
and all of its time derivatives for all values of $\vec{x}\neq 0$.
The only allowed singularity of the full solution is point-like, and located at the origin $t=0=\vec{x}$.

As was pointed out in \cite{Son}, the construction of the saddle-point solution above 
has a more natural implementation in terms of the complex-valued time coordinate.
In Minkowski space-time $x^\mu=(t,\vec{x})$ the solution $h(x)$ contains a point-like singularity at the origin $x=0$
arising from the delta-function source term in \eqref{sing_eqn}, and is regular everywhere else. In the 
Euclidean space-time, $(\tau,\vec{x})$, however, the solution will in general be singular on a hypersurface
$\tau = \tau_0(\vec{x})$. For a particularly simple case of the uniform in space solution
\eqref{clas_sol2}, the singularity surface is $\tau_0=\tau_\infty$ which is an $\vec{x}$-independent constant as we 
have already seen. This solution describes the generating functional of tree-level amplitudes on $n$-particle mass thresholds.
In the more non-trivial settings, specifically in the case of solutions relevant for the processes away from the
multi-particle thresholds, or beyond the tree-level, or both, the relevant fields do depend on the spacial variable and
as the result, the singularity surface $\tau = \tau_0(\vec{x})$ is an ${\cal O}(3)$ symmetric function of the spacial variable.
Consider the singularity surface $\tau_0(\vec{x})$ of the form shown in the Figure (1a). It is a local 
deformation of the flat singularity domain wall at $\tau_\infty$ with the single maximum touching the origin $(\tau, \vec{x})=0$.
As such, in Minkowski space the singularity is point-like at $t=i\tau=0$ and $\vec{x}=0$ as required.

 \begin{figure}
\begin{center}
\hskip-1.5cm \includegraphics[width=1.08\textwidth]{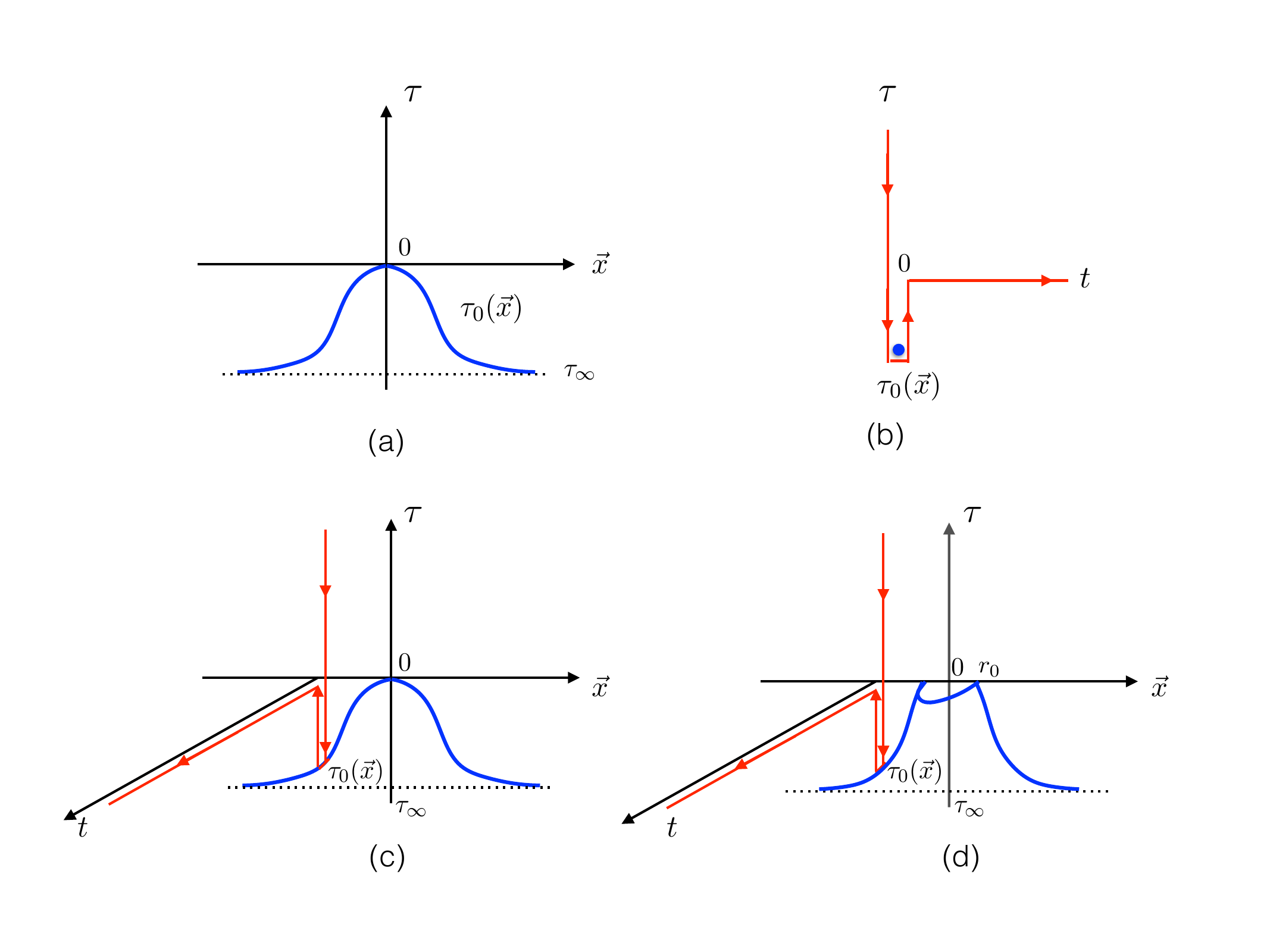}
\end{center}
\vskip-.6cm
\caption{{\bf Figure~(1a)} shows the singularity surface $\tau=\tau_0 (\vec{x})$ of the field configuration $h$ 
on the imaginary-time hyper-plane $(\tau,\vec{x})$. The tip of the singularity surface is located at $\tau=0$ so that 
in Minkowski space-time the solution is singular only at a single point taken to be the origin $(t, \vec{x})=(0,\vec{0})$. 
Away from the local maximum, the singular domain-wall 
$\tau_0 (\vec{x})$ approaches the constant space-independent value $\tau_\infty$.
{\bf Figure~(1b)} shows the time-evolution contour on the complex time plane. The two turning points are the location of the
singularity surface, $\tau_0 (\vec{x})$, and the origin, $\tau=0=t$, after which evolution to the final state proceeds along 
the real $t$~axis.
{\bf Figure~(1c)} shows the same time-evolution contour at a fixed value of $\vec{x}$ 
along with the singularity surface of the classical field in the complex-time--space coordinate system $(t,\tau,\vec{x})$.
{\bf Figure~(1d)} is the same as (c), but the singular domain wall $\tau_0 (\vec{x})$ is folded into the real time direction
for $|x| < r_0$ where $r_0$ is the critical radius of the domain wall bubble.
}
\label{fig:4}
\end{figure}

Thus by extending the real time variable into the complex plane we have extended the point-like singularity of the solution
to the singularity hypersurface $\tau = \tau_0(\vec{x})$ or the singular domain wall. The next step is to define the 
time evolution contour of the solution in the complex plane from the initial to the final time boundaries.
It is shown in the Figure (1b). 
At early times the solution evolves along the imaginary-time axis from the initial time boundary
at $\tau=+\infty$ down to the singularity surface of the solution at $\tau_0$. The contour then encircles the singularity
$\tau_0$ at each fixed value of $\vec{x}$ and evolves upwards along the imaginary-time axis to $\tau=0$. From there
on the third segment the contour evolves along the real-time axis from $t=0$ to the final-time boundary at $t\to +\infty$.
The Figure (1c) shows this contour in the $(t,\tau,\vec{x})$ coordinates along with the singularity surface 
of the solution at $\tau=\tau_0(\vec{x})$.

\bigskip

We now return to the two branches of the solution $h_1(\tau,\vec{x})$ and $h_2(t,\vec{x})$ introduced in the beginning 
of this section, but now defined along the time evolution contour in Fig.~\ref{fig:4}.
Both these field configurations are finite regular classical solutions on the subspaces defined by
$\tau > \tau_0(\vec{x})$ for $h_1(\tau,\vec{x})$, and by $t>0$ for $h_2(t,\vec{x})$. They satisfy the 
boundary conditions ({\it cf.} \eqref{eq:limin}-\eqref{eq:limf} and recall the substitution $t=i\tau$ for the initial time asymptotics),
\begin{eqnarray}
\lim_{\tau\to + \infty}\, h_1(\tau,\vec{x})  \,-\, v &=& 0 \,
\\
\lim_{t\to + \infty}\,h_2(t,\vec{x}) \,-\, v  &=& 
\int \frac{d^3k}{(2\pi)^{3/2}} \frac{1}{\sqrt{2\omega_{\bf k}}}\left( b_{\bf k}\,e^{\omega_{\bf k}T-\theta}
\, e^{-ik_\mu x^\mu}\,+\, b^\dagger_{\bf k}\, e^{ik_\mu x^\mu}\right)\,.
\end{eqnarray}
The Euclidean action of the complete solution $h(x)$ along our complex-time contour can be straightforwardly 
represented as the appropriate action integrals of the solutions $h_1(\tau,\vec{x})$ and $h_2(t,\vec{x})$ 
on the parts of the contour,
\[
S_{\rm Eucl}\,=\,\int  d^3 x \left[ -\int_{+\infty}^{\tau_0(\vec{x})} d\tau \, {\cal L}_{\rm Eucl}(h_1) \,-\,
\int_{\tau_0(\vec{x})}^0d\tau \, {\cal L}_{\rm Eucl}(h_2) \,-\,i 
 \int_{0}^\infty dt \, {\cal L}(h_2) \right]\,,
 \label{eq:Se}
\]
where we used the standard notation $ {\cal L}_{\rm Eucl}(h)\,=\, \frac{1}{2}\left(\partial_\mu h\right)^2 \,+\, V(h)$,
$S_{\rm Eucl}\,=\,- i\,  S$ and recalled that  $\tau=-t_{\rm Eucl}$ (which explains the minus signs in the first two terms). 


Up to this point we have not attempted to impose any matching conditions on the
two trial functions, $h_1(\tau,\vec{x})$ and $h_2(\tau,\vec{x})$,
at the singularity. Without the matching, the two individual components are some easy-to-obtain classical solutions 
with the correct boundary conditions at the initial $\tau \to \infty$ and final $t\to \infty$ times, but they do not solve the 
required boundary value problem. First one can imagine adjusting the coefficients $b_{\bf k}$ to match the two profiles
on a certain candidate surface $A$, so we set $h_1\,=\, h_2 \,=\, \Phi_0$ on $A$. We assume a regularisation procedure 
which keeps $\Phi_0$ finite at intermediate stages of the calculation to avoid infinities. The approximation to the 
true saddle-point is still very crude as the derivatives normal to the surface do not match, and the 
matching of $h_1$ and $h_2$ has a cusp on the entire surface $A$,
\[ \partial_n (h_1\,-\,h_2)\,=\, J(A) \,.
\]
This defines a function $J(A)$ supported on the surface of $A$.
It then follows that the field configuration $h(x)$ obtained from this matching satisfies 
the equation 
\[ \frac{\partial S_{\rm Eucl}}{\partial h}\,=\, J(x) \,, \quad {\rm where} \quad
J(x) \,=\, \int d^3A \, J(A) \, \delta^{(4)}(x-x(A))\,.
\]
This is the classical equation with a source $J(x)$ rather than
$j \delta^{(4)}(x)$ appearing in the equation \eqref{sing_eqn} we are meant to be solving. 
The important point emphasised in \cite{Son} is that we can repair this by varying the
shape of the candidate surface $A$, and consider a family of Euclidean actions 
of the form \eqref{eq:Se} each computed using a particular surface $A$. Then it is easy to show
that once the action $S_{\rm Eucl}[h]$ has been extremized with respect to $A$, the source $J(x)$ 
corresponding to the extremal surface becomes of the required $\delta$-function form, $j \delta^{(4)}(x)$.

\bigskip

This concludes our review of the semiclassical method of Son \cite{Son}. 
To summarise our main conclusion in this section,
it was shown that the required saddle-point solution to the multi-particle boundary-value problem  
can be obtained by extremizing the real part of the Euclidean action over all singularity surfaces $\tau=\tau_0(\vec{x})$
containing the point $t=0=\vec{x}$.

There are two equivalent formulations of the problem. One either finds the required solution with 
the point-like singularity at the origin by varying the Fourier coefficients of the solution asymptotics
at $t \to \pm \infty$, or alternatively, one extremizes the classical action by varying the singularity surfaces 
of the solutions in complex time. The second method will be particularly well suited for using the thin-wall 
approximation in the following section.
This will allow us to compute the dominant contribution to $W(E,n)$ in the limit $\lambda n\to \infty.$


\medskip
\section{Thin wall critical bubbles}
\label{sec:4}
\medskip

The main goal of this section is to use the semiclassical method described above
 to carry out a novel computation of the multi-particle rates ${\cal R}_n(E)=\,e^{W(E,n)}$
in the large $\lambda n$ limit.
This  involves the higher-loop quantum effects, and in order to correctly address them we first need to
normalise on the known tree-level high multiplicity results.
Our starting point is the function $W$ in \eqref{eq:Wdef2} appearing in the exponent of the rate which is 
evaluated on the saddle-point solution. In terms of the Euclidean action it is given by,
\[ W(E,n; \lambda) \,=\, \frac{1}{\lambda}\, {\cal F}(\lambda n, \varepsilon)\,=\,
ET \,-\, n\theta \,-\, 2S_{\rm Eucl}[h]\,.
\label{eq:Wdef2}
\]
At tree-level, the function $W$ is of the form,
\[
W(E,n; \lambda)^{\rm tree}  \,=\,\frac{\lambda n}{\lambda}\left( f_0(\lambda n)\,+\, f(\varepsilon) \right)\,,
\label{eq:Wtree}
\]
and its dependence on $\lambda n$ and on the average kinetic anergy per particle per mass, $\varepsilon$,
is in terms of two individual functions of each argument, $f_0(\lambda n)$ and $f(\varepsilon)$.
These functions are known,
\begin{eqnarray}
\label{f0SSB}
f_0(\lambda n)&=&  \log\left(\frac{\lambda n}{4}\right) -1\,, 
\\
\label{feSSB}
f(\varepsilon)|_{\varepsilon\to 0}&\to& f(\varepsilon)_{\rm asympt}\,=\, 
\frac{3}{2}\left(\log\left(\frac{\varepsilon}{3\pi}\right) +1\right) -\frac{25}{12}\,\varepsilon\,,
\end{eqnarray}
where the expression \eqref{feSSB} is valid in the non-relativistic limit $\varepsilon \ll 1$
near the multi-particle mass-threshold.
These tree-level results \eqref{eq:Wtree}-\eqref{feSSB} were computed using both types of methods: the resummation of
Feynman diagrams based on solving the recursion relations and integrating over the phase-space in \cite{LRST,VVK2}, 
and also from using the semiclassical approach \cite{Son} directly. The apparent agreement between the two methods
provides a useful consistency check
on the semiclassical formalism.

The above result has also been generalised to  the general tree-level kinematics. 
In particular, at tree-level the function $f_0(\lambda n)$ is fully determined, but the second function,
 $f(\varepsilon)$, characterising the energy-dependence of the final state, is determined by Eq.~\eqref{feSSB} 
 only at small $\varepsilon$, i.e. near the multi-particle threshold. This point was addressed in Ref.~\cite{VVK3}
 where the function $f(\varepsilon)$ was computed numerically in the entire range  $0\le \varepsilon < \infty$. 
 
 It is also known how to add the leading-order loop corrections to the tree level expressions in the $\lambda n \ll 1$ limit.
 This has been achieved in Ref.~\cite{LRST}
 by resumming the one-loop correction to the amplitude on the multi-particle mass threshold 
 originally computed in Refs.~\cite{Voloshin:1992nu,Smith:1992rq}. The same
 result was also reproduced using the semiclassical method \cite{Son}, once again providing a valuable justification of this
 approach.
This results in the modified expression for $f_0$, 
\[
f_0(\lambda n)^{\rm 1-loop} \,=\, \log\left(\frac{\lambda n}{4}\right) -1 \,+\, \sqrt{3}\, \frac {\lambda n}{4\pi}\,.
\label{eqnl}
\]

\bigskip

To  determine whether the bare\footnote{By the {\it bare} rate we mean the rate
with an external i.e. non-dynamical initial state given by ${\cal O}|0\rangle$. The Higgsplosion effect of
\cite{Higgsplosion} is the result of the exponentially growing {\it bare} rate ${\cal R}_n(E)$. As explained in 
\cite{Higgsplosion} the physical cross-sections involve instead the rates corrected by the resummed i.e. 
dynamical propagator of the initial state; these physical cross-sections do not explode and are consistent with 
the unitarity of the theory. This was called  the Higgspersion effect in Ref.~\cite{Higgsplosion}.}
 multi-particle rate ${\cal R}_n(E)$ 
defined in \eqref{eq:RnE1} and \eqref{eq:ReW} can become exponentially large above a certain critical 
particle number $n$ and lead to a realisation
Higgsplosion  \cite{Higgsplosion} in a given theory, we need to be able to address the large $\lambda n$ limit.
Up to now the loop effects were only computed in the opposite regime of small $\lambda n$ in \eqref{eqnl}.

In the following we will address the large-$n$ limit with
the value of the combination $\lambda n$ taken to be large, $\lambda n \gg 1$, while the average particle energy is kept
non-relativistic, $\varepsilon \ll 1$. This selects the regime of multiplicities $n$ approaching their maximal  
values allowed by the fixed energy kinematics, $n \sim n_{\rm max}=E/M_h$ where $\varepsilon \sim 0$ 
and the final state particles are non-relativistic.

\bigskip

The tree-level result \eqref{eq:Wtree}-\eqref{feSSB} in the non-relativistic limit $\varepsilon \to 0$ arises 
in the semiclassical calculation from the uniform in space saddle-point solution \eqref{clas_sol2}.
As we have already discussed, this solution corresponds to a singular domain wall located at a constant 
value of $\tau$, so that the singularity surface does not depend on $\vec{x}$. In the large $\lambda n$ limit
one should be able to similarly write down a singular field configuration that serves as the saddle-point of the path integral 
representation of the multi-particle rate ${\cal R}_n$. The singularity surface of this configuration is however locally deformed 
by the source at $x=0$, while at large values of $|\vec{x}|$ the singularity surface $\tau_0(\vec{x})$ rapidly approaches the 
constant value $\tau_0(\vec{x})\to \tau_\infty$.

In both of these cases we need to fix the translational symmetry of the solutions by locating the singularity surfaces 
in such a way that its local maximum is at 
the point $\tau=0=\vec{x}$. 
This implies that the flat domain wall used for calculating the tree-level amplitudes on $n$-particle thresholds 
should be located at $\tau=0$. At the same time, the $n$-particle amplitudes in the large $\lambda n$ limit
arise from the feld configuration with the 
singularity surface located at $\tau_0(\vec{x})$, as described above. Each of these amplitudes are determined by the 
$ z(\tau)^n \sim e^{-n M_h \tau}$
term in the corresponding Taylor expansion of the field configurations, as in \eqref{gen-funh2}.
Now the difference between the singularity surface located at $\tau=0$ and at $\tau=\tau_0 (\vec{x})\to \tau_\infty$
rescales the ${\cal A}_{1\to n}$ amplitude on the $n$-particle threshold by a multiplicative factor of
$e^{-nM_h \tau_\infty}$.

This is not all. We still need to determine the shape of the curved singularity surface $\tau=\tau_0(\vec{x})$ 
by requiring that it extremizes the Euclidean action on the corresponding singular solution, as was discussed in the
previous section. Hence we need to add to the exponent of the rate the factor  
$-\, 2 S_{\rm Eucl} [\tau_0(x)] \,+\, 2 S_{\rm Eucl} [0]$ where the last term removes the 
contribution of the flat wall (already accounted in the tree-level result).
These simple qualitative arguments lead to the following form of the $W$ function in the large $\lambda n$ limit
(note the factors of 2 arising from squaring the amplitudes),
\[ W(E,n; \lambda) \,=\, W(E,n; \lambda)^{\rm tree}\, -\, 2n M_h \tau_\infty\,-\, \left(2 S_{\rm Eucl} [\tau_0(x)]
-\, 2 S_{\rm Eucl} [0] \right)
\,,
\label{eq:WlnL}
\]
where by $S_{\rm Eucl}$ we mean the Real part of the Euclidean action (or equivalently the 
Imaginary part of the Minkowski action).
The expression \eqref{eq:WlnL} is supposed to be valid in the double-scaling large-$n$ limit \eqref{eq:limit}
where the two scaling quantities $\lambda n$ and
$\varepsilon$ are such that $\lambda n \gg 1$ and $\varepsilon \ll 1.$
The singularity surface $\tau_0(x)$, its asymptotics $\tau_\infty$ and the Euclidean action itself 
will now need to be determined as functions of $\lambda$, $\lambda n$ and $\varepsilon$
by extremizing $S_{\rm Eucl}$ as the functional of  $\tau_0(x)$.

Before we proceed with finding the saddle-point singularity surface for the action, 
it is worthwhile to note that the same conclusion
was also derived in the Section 4.1 of Ref.~\cite{Son} using a more technical direct approach based on solving
the boundary-value problem using a deformation of the flat-wall solution in the form
\[
h(\tau,\vec{x})\,=\, v\, \left(\frac{ 1\,+\,e^{-M_h (\tau-\tau_\infty)}}
{1\,-\, e^{-M_h (\tau-\tau_\infty)}}\right)
\,+\, \delta h(\tau,\vec{x})
\]
with the support on the singular surface $\tau=\tau_0(x)$.

\bigskip

The problem of finding the large-$\lambda n$ correction
$\frac{1}{\lambda} \, g(\lambda n)$  to the $W$ function has a simple geometric interpretation.
We need to maximise the expression
\begin{eqnarray}
\frac{1}{2\lambda} \, g(\lambda n) &=& - n M_h \tau_\infty \,-\,  {\rm Re} (S_{\rm Eucl} [\tau_0(x)]-S_{\rm Eucl} [0]) \nonumber\\
 &=& n M_h |\tau_\infty| \,-\,   {\rm Re} (S_{\rm Eucl} [\tau_0(x)]-S_{\rm Eucl} [0])\,,
\label{eq:gcorr}
\end{eqnarray}
where we have used the fact that $\tau_\infty$ is negative and hence the first term on the right hand side of \eqref{eq:gcorr}
 is positive-valued.
This extremization problem  corresponds to finding the shape of the membrane with the surface tension
dictated by the action $S_{\rm Eucl}$ and located at the position $\tau_\infty$ which is pulled at its centre by a constant force.

The main idea on which our calculation will be based is the geometrical interpretation of the saddle-point 
field configuration as a domain wall solution separating the vacua with  different VEVs $h \to \pm v$ on the different
sides of the wall. Our scalar theory with the spontaneous symmetry breaking in \eqref{eq:L} clearly supports
such field configurations.\footnote{We expect that a similar approach will also work in the full weak sector of the 
Standard Model where the simplified description \eqref{eq:L} applies to the single scalar degree of freedom in the
unitary gauge. We imagine first selecting  the processes with the multiple production of scalars only in the 
final state. The SM vector bosons and fermions would also contribute here as the virtual states in the loops,
 along with the self-interactions of the scalars.
The calculation in the present paper  will account only for the scalar self-interaction effects in the large $\lambda n$ limit,
while the investigation of the role and size of quantum effects due to virtual vectors and fermions is left for future work.}
The solution is singular on the surface of the wall, and the wall thickness is $\sim 1/M_h$. The effect of the `force'
$nM_h$ applied to the domain wall locally pulls upwards  the centre of the wall and gives it a profile $\tau_0(\vec{x})$
depicted in Fig.~\ref{fig:4}. To find the equilibrium position of the domain wall one needs to find an extremum of 
the expression in \eqref{eq:gcorr}. When computing the Euclidean action on the solution characterised by the
domain wall at $\tau_0(\vec{x})$, it will be represented by the action of a thin-wall bubble. The shape of the bubble will
be straightforward to determine by extremizing the action in the thin-wall approximation, and the validity of this approximation 
will be shown to be justified in the limit $\lambda n \to \infty$. Our implementation of this set-up 
will follow closely the construction of Gorsky and Voloshin in Ref.~\cite{GV}.

The Euclidean action computed along the complex time evolution contour shown in Figure (1b) 
is given by the sum of three contributions, each of them corresponding to one of the three segments of the 
integration contour. This structure  $S_{\rm Eucl}= S_{\rm Eucl}^{(I)} + S_{\rm Eucl}^{(II)} -i S^{(III)}$
is manifest in the expression on the right hand side of \eqref{eq:Se}. 
But only the first two segments contribute to the Real part of $S_{\rm Eucl}$ appearing in the rate in \eqref{eq:gcorr}.

The real part of the action \eqref{eq:Se} computed on the field $h(x)$ which is characterised by the 
surface of singularities $\tau = \tau_0(\vec{x})$ can be written as an integral on the singularity surface 
in the thin-wall approximation. This is equivalent to stating that
 the action is equal to the surface tension of the domain wall $\mu$ times the area $A$.
We have,
\[ S_{\rm Eucl}[ \tau_0(\vec{x})] \,=\, \int_{\tau_\infty}^0 d\tau \, L (r,\dot r)\,=\,
\int_{\tau_\infty}^0 d\tau \,4\pi \mu \, r^2 \sqrt{1+\dot r^2}
\,,
\label{eq_thinw}
\]
where $r=|\vec{x}|$ and  $\dot r= dr/d \tau$. The integral depends on the choice of the domain wall surface 
$\tau_0(\vec{x})$
implicitly via dependence on $\tau$ of $r(\tau)$ and $\dot r(\tau)$ which are computed on the domain wall.
For the surface tension we have \cite{GV},
\[
\mu\,=\, \int_{-\infty}^{\infty} d\tau \left( \frac{1}{2} \left(\frac{d h_{\rm cl}}{d\tau}\right)^2\,+\,
 \frac{\lambda}{4} \left( h_{\rm cl}^2 - v^2\right)^2 \right) \,=\,  \frac{M_h^3}{3 \lambda}\,
 \label{eq:mu}
\]
where integral in \eqref{eq:mu} is computed on the flat domain wall solution \eqref{clas_sol2},
\[
h_{\rm cl}(\tau)\,=\, v\, \left(\frac{ 1\,+\,e^{-M_h (\tau-\tau_\infty)}}
{1\,-\, e^{-M_h (\tau+\tau_\infty)}}\right)
\label{eq:mu2}
\]
with the
argument $\tau$ in the equation above further shifted by $i \theta$ with a constant real-valued parameter 
$\theta$. The physical reason for this shift is that, as we will see below, the singular domain wall will be folded into the real-time direction
for $x$ less than the critical radius $r_0$ of the bubble as indicated in Fig. (1d). Hence the integration contour 
along the 
$\tau$-direction is actually sifted along the real $t$ axis.\footnote{I thank Joerg Jaeckel for a useful discussion of this point.}
This also ensures that the integration contour in \eqref{eq:mu} is shifted away 
from the pole of  \eqref{eq:mu2} at $\tau=0$. The field
configuration in \eqref{eq:mu2} is periodic with the period $2i\pi/M_h$, and the poles are located at 
$M_h (\tau-\tau_\infty)=0,\pm 2i\pi,\pm 4 i\pi,$ etc. It is easy to see (by closing the integration contour into a loop that does not 
surround any poles in $\tau$)
that any non-vanishing value of $\theta$ in the range $0< M_h\theta<2\pi$
would give the same expression $ \frac{M_h^3}{3 \lambda}$ for the integral  \eqref{eq:mu}. For simplicity
we can then choose $M_h \theta=\pi$ and evaluate the surface tension integral in \eqref{eq:mu} on the regular real-valued configuration,
\[
h_{\rm cl}(\tau+i\pi/M_h)\,=\, v\, \left(\frac{ 1\,-\,e^{-M_h \tau}}
{1\,+\, e^{-M_h \tau}}\right)\,,
\label{eq:mu3}
\]
which leads to the expression for the surface tension on the right hand side of \eqref{eq:mu}.
\bigskip
 
The contribution to the function $-\frac{1}{2\lambda} \, g(\lambda n)$ computed on its saddle-point 
can be recast as follows,
\begin{eqnarray}
nM_h \tau_\infty \,+\,  S_{\rm Eucl}  &=& \left(nM_h -E\right) \tau_\infty \,+\, 
 \int_{\tau_\infty}^0 d\tau \left( L -E\right)      \label{eq:1}    \\
 &=& \left(nM_h -E\right) \tau_\infty \,+\, 
 \int_{\tau_\infty}^0 d\tau \left( L -H\right)         \label{eq:2}   \\
  &=& \left(nM_h -E\right) \tau_\infty \,+\, \int_R^{r_0} p(E,r) \, dr\,.   \label{eq:3}
\end{eqnarray}
On the first line we have subtracted and added the constant $E$ which we take to be the energy 
of the domain wall. The extremum of the overall expression above is achieved by extremizing the action as well
as differentiating with respect to $\tau_\infty$. The former condition implies that the surface of the wall satisfies the
Euler-Lagrange equations of motion corresponding to the Lagrangian $L$. On these solutions their energy is an 
integral of motion and is equal to the Hamiltonian $H$. On the second line \eqref{eq:2} we have traded $E$ for $H$
in the integral. 
The Hamiltonian is defined in terms of the usual Legendre transformation of the Lagrangian function
$L (r,\dot r) =\, 4\pi \mu \, r^2 \sqrt{1+\dot r^2}$,
\[
H(p,r)\,=\, L (r,\dot r) \,-\, p\, \dot r\,,
\label{Hdef}
\]
where the momentum $p$, which is the variable conjugate to the coordinate $r$, is defined via (the signs 
correspond to the Euclidean formulation):
\[
p\,=\, \frac{\partial L (r,\dot r) }{\partial \dot r}\,\,\,=\, 4\pi\, \mu \frac{r^2 \dot r}{\sqrt{1+\dot r^2}}
\label{eq:defp1}
\]
Thus we see that the integral $\int (L-H) d\tau$ on the right hand side of \eqref{eq:2} 
is equivalent to the integral $\int  p \, dr$ appearing in \eqref{eq:3}. 
We further note that the conjugate momentum appearing on the right hand side
of Eq.~\eqref{eq:defp1} is given by the negative semi-definite expression, $p(\tau) \le 0$.
This is because the radius of the solution, $r(\tau)$, is a monotonically decreasing function of its argument as $\tau$ varies from the lower limit 
$\tau_\infty$ where $r(\tau_\infty)=\infty$ to a higher value $\tau_0 >\tau_\infty$  where the radius $r(\tau_0)=r_0$ is finite. Hence the time derivative 
$\dot r$ is $ \le 0$ and so is the conjugate momentum function $p(\tau) \propto \dot r$ in \eqref{eq:defp1}. 
Using this we can swap the integration limits in the expression in \eqref{eq:3} and write,
\[
nM_h \tau_\infty \,+\,  S_{\rm Eucl} \,=\,
\left(nM_h -E\right) \tau_\infty \,+\, \int^R_{r_0} |p(E,r)| \, dr\,.   \label{eq:3ag}
\]
The variation of \eqref{eq:3ag} with respect to $\tau_\infty$ imposes the constraint $E=nM_h$,
which can be understood as the fact that in the $n$-particle threshold limit, the energy of the field is
the energy in the final state  which is given by $nM_h$ for $\varepsilon=0$. 
Thus we have for the extremum,
\[ E \,=\, nM_h\,, \qquad 
E \tau_\infty \,+\,  S_{\rm Eucl}  \,=\,  \int^R_{r_0} |p(E,r)| \, dr\,,
\label{eq:pdrN}
\]
where $|p|$ is the absolute value of the momentum and, as we will see momentarily, for the classical solution
of energy $E$, it can be written in the form,
\[
|p(E,r)| \,=\, 4\pi \, \mu \, \sqrt{r^4- \left(\frac{E}{4\pi\mu}\right)^2}\,.
\label{eq:conjp}
\] 
The lower bound of the integral in \eqref{eq:pdrN} is cut-off at the critical radius $r_0$,
\[
r_0^2 \,=\, \frac{E}{4\pi\mu} \,,
\label{r0def}
\]
which is the smallest possible radius of the bubble for which the conjugate momentum in 
\eqref{eq:conjp} is well-defined. The upper bound of the integral \eqref{eq:pdrN} is at $R\gg 1$ 
which will be ultimately taken to infinity.
To derive the expressions on the right hand sides of \eqref{eq:pdrN} and \eqref{eq:conjp}, it is useful to 
re-write \eqref{Hdef} in the form,
\[E\,=\, 4\pi \mu \, r^2 \sqrt{1+\dot r^2} \,-\, 4\pi\, \mu \frac{r^2 \dot r}{\sqrt{1+\dot r^2}}\,=\,
4\pi\, \mu \frac{r^2}{\sqrt{1+\dot r^2}}\,,
\]
and then compute the combination using the above expression and \eqref{eq:defp1},
\[
E^2\,+\, p^2\,=\, \left(4\pi \mu \,r^2 \right)^2 \left( \frac{1}{1+\dot r^2}\,+\, \frac{\dot r^2}{1+\dot r^2}\right)\,=\,
 \left(4\pi \mu \,r^2 \right)^2\,.
 \]
With this (and selecting the minus sign for the momentum in accordance to the monotonically decreasing $r(\tau)$) we have,
\[ 
p \,=\,-\,4\pi \, \mu \, \sqrt{r^4- r_0^4}\,,
\label{eq:conjp2}
\] 
which of course is equivalent to \eqref{eq:conjp}.

\bigskip

What is the meaning of the critical radius $r_0$ in \eqref{eq:conjp2} and \eqref{r0def}?
It is the minimal value of the radius of the bubble, $r(\tau) \ge r_0$, formed by the membrane pulled 
by the force $\sim E$. The radius $r_0$ is the turning point of the solution, $\dot r_0 =0$; if one tries to go to smaller values of the radius,
the conjugate momentum $p(r)$ becomes complex below $r=r_0$. What this implies for our construction is that
the surface of singularities $\tau=\tau_0(\vec{x})$ gets folded into the real-time axis for $r\le r_0$. This is 
sketched in the Figure (1d). For all practical purposes this simply implies that
the integral in the action in \eqref{eq:pdrN} has the lower limit at $r=r_0$.

\bigskip

We can now evaluate the  correction $\frac{1}{\lambda} \, g(\lambda n)$ to the $W^{\rm tree}$ function
in the large $\lambda n$ limit. In order to proceed with this task, note that we still have to subtract the
contribution to the action of the flat domain wall solution. 
Hence we have in total 
\[ -\frac{1}{2\lambda} \, g(\lambda n) \,=\, 
E \tau_\infty \,+\, S_{\rm Eucl} [\tau_0]\,-\,  S_{\rm Eucl}[0]  \,=\, 
 \int^R_{r_0} |p(E)| \, dr\,-\, \int^R_0 |p(E=0)| \, dr 
\,,
\label{eq:pdrtot}
\]
where $E=nM_h$ as before and $|p(E)|$ is given by \eqref{eq:conjp}.
This is evaluated as follows. We use the trick of \cite{GV} to introduce the identity $1=\int dE \, d/(dE)$
and thus re-write the right hand side of \eqref{eq:pdrtot} as follows,
\begin{eqnarray}
\int_0^E dE \left( \frac{d}{dE}\, \int^R_{r_0} |p(E)| \, dr \right) &=&
-\int_0^E dE \, \frac{E}{4 \pi \mu}  \,\int^R_{r_0}\frac{d r}{\sqrt{r^4-r_0^4}}\\
&=& -\int_0^E dE \sqrt{E}\, \frac{1}{\sqrt{4\pi \mu}} \, \int_1^\infty \frac{dx}{\sqrt{x^4-1}}\, \,=\,
-\, \frac{E^{3/2}}{\sqrt{\mu}}\, \frac{1}{3}\, \frac{\Gamma(5/4)}{\Gamma(3/4)}\,. \nonumber
\end{eqnarray}

In summary, our final result for the quantum correction to the exponent of multi-particle rate in the 
large $\lambda n$ limit is given by
\[
\frac{1}{\lambda} \, g(\lambda n) \,:=\, \Delta W(E,n;\lambda) \,=\,\, \frac{1}{\lambda} \, (\lambda n)^{3/2}\, \frac{2}{\sqrt{3}}\,
\frac{\Gamma(5/4)}{\Gamma(3/4)}\,\simeq\, \, 0.854\,  n \sqrt{\lambda n} \,.
\label{eq:gln}
\]
We note that this expression is positive-valued, that it grows in the limit of $\lambda n \to \infty$, and that it has
the correct scaling properties for the semiclassical result, i.e. it is of the form $1/\lambda$ times a function of $\lambda n$.

Numerically, our result agrees with the expression derived in Ref.~\cite{GV} for the case of $d=3$ spacial dimensions.
It also follows that the thin-wall approximation is fully justified in our $\lambda n \gg 1$ limit.
The thin-wall regime corresponds to the radius of the bubble being much greater than the thickness of the wall,
$r \gg 1/M_h$. In our case the radius is always greater than the critical radius,
\[
r\,M_h\, \ge\, r_0\,M_h\,=\, M_h\,\left( \frac{E}{4\pi \mu}\right)^{1/2}\,\sim\, \left( \frac{\lambda\, E}{M_h}\right)^{1/2} =\,
 \sqrt{\lambda n}\,\,\gg\, 1\,.
\]

 \bigskip
\section{Discussion of the results}
\label{sec:5}
\bigskip

We have computed the quantum contributions to the exponent of the multi-particle 
production rate that are dominant in the high-particle-number $\lambda n \to \infty$ 
limit in the kinematics of near-maximal $n$ where the final state particles are produced near their mass thresholds.
This corresponds to the limit 
\[ \lambda \to 0\,, \quad n\to \infty\,, \quad {\rm with}\quad
\lambda n = {\rm fixed} \gg 1 \,, \quad \varepsilon ={\rm fixed} \ll 1 \,.
\label{eq:limit2}
\]
The resulting quantum-effects-corrected 
multi-particle production rate at energy $E$ is one of the main results of this paper and it 
is given by a characteristic exponential-form
representation in the limit \eqref{eq:limit2}, obtained by combining the previously known tree-level contribution \eqref{eq:Wtree}
with our new result \eqref{eq:gln}. We have,
\[
{\cal R}_n(E)\,= \, e^{W(E,n)}\,=\, 
\exp \left[ \frac{\lambda n}{\lambda}\, \left( 
\log \frac{\lambda n}{4} \,+\, 0.85\, \sqrt{\lambda n}\,-\,1\,+\,\frac{3}{2}\left(\log \frac{\varepsilon}{3\pi} +1 \right)
\, -\,\frac{25}{12}\,\varepsilon 
\right)\right] 
\label{eq:Rnp2}
\]
This expression for the multi-particle rates was used in Ref.~\cite{Higgsplosion} to motivate and
illustrate the Higgsplosion mechanism. 
The expression \eqref{eq:Rnp2} was derived in the near-threshold limit where
the parameter $\varepsilon$ is treated as a fixed number much smaller than one. 
The energy in the initial state and the final state multiplicity are related linearly via
\[ E/M_h \,=\, (1 + \varepsilon) \, n\,,
\]
and thus for any fixed non-vanishing value of $\varepsilon$, one can raise the energy to 
achieve any desired large value of $n$ and consequentially a large $\sqrt{\lambda n}$.
Clearly, at the strictly vanishing value of $\varepsilon$, the phase-space volume is zero and 
the entire rate \eqref{eq:Rnp2} vanishes. Then by increasing $\varepsilon$ to a positive but still small values, the rate increases.
The competition is between the negative $\log \varepsilon$ term and the positive $\sqrt{\lambda n}$ term 
in \eqref{eq:Rnp2}, and there is always a range of sufficiently high multiplicities where $\sqrt{\lambda n}$
overtakes the logarithmic term $\log \varepsilon$ for any fixed (however small) value of $\varepsilon$.
This leads to the exponentially growing multi-particle rates above a certain critical energy, which in the case
described by the expression in \eqref{eq:Rnp2} is in the regime of $E_c \sim 200 M_h$.

 \begin{figure*}[t]
\begin{center}
\begin{tabular}{c}
\hskip-.6cm \includegraphics[width=0.47\textwidth]{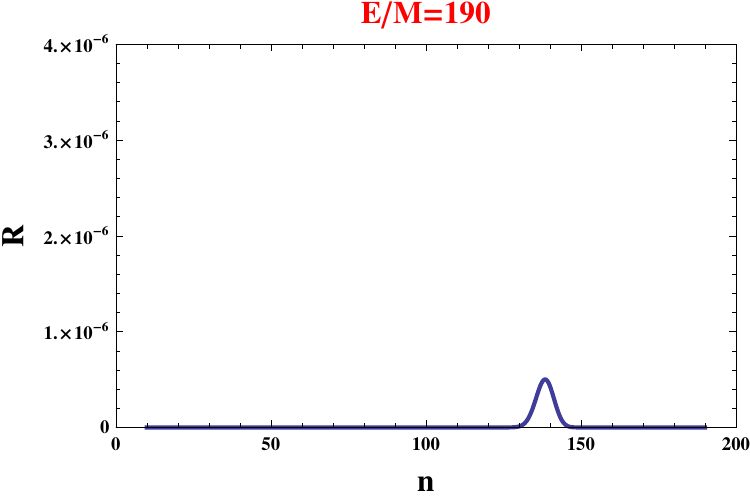}\quad
\includegraphics[width=0.45\textwidth]{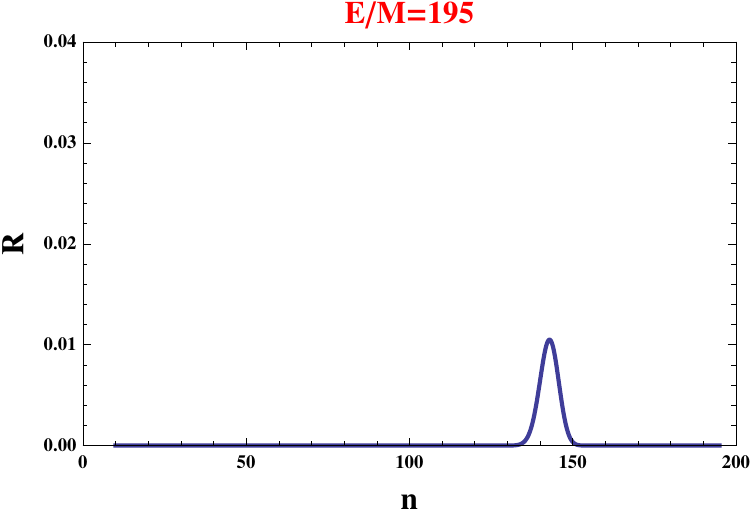}
\\
\\
\includegraphics[width=0.45\textwidth]{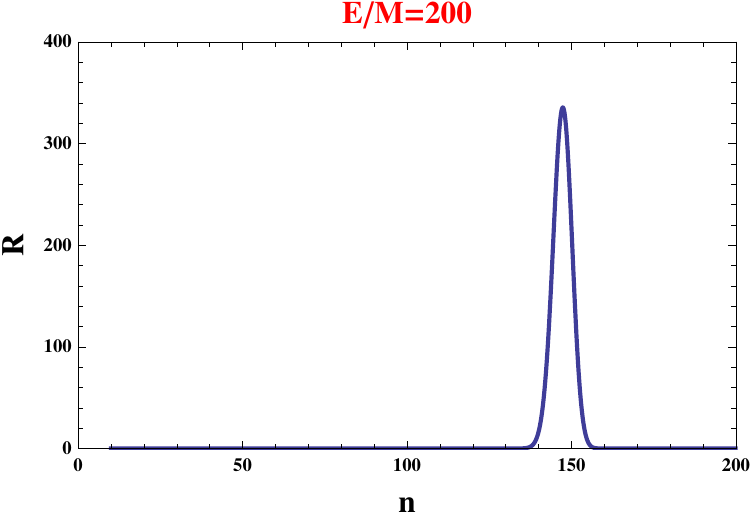}\quad
\includegraphics[width=0.45\textwidth]{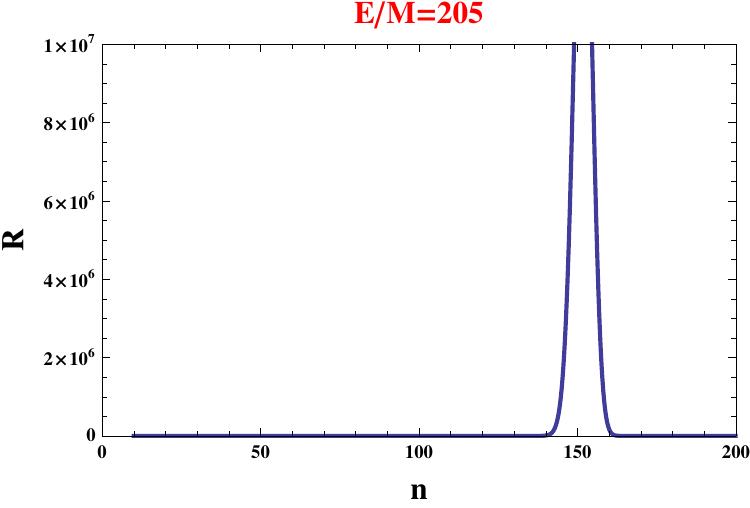}
\end{tabular}
\end{center}
\vskip-.6cm
\caption{Multi-particle decay rates Eq.~\eqref{eq:Rnp2}
of a highly-energetic single-particle state into 
$n$ scalars plotted as function of $n$. The four sub-figures show the energy $E$ fixed at 
190$M_h$, 195$M_h$, 200$M_h$ and 205$M_h$ and we used $\lambda=1/8$. There is a sharp exponential dependence of the peak rate
on the energy. The peak multiplicities $n \sim 150$ in these examples are not far from the maximally allowed values 
at the edge of the phase space $n_{\rm max} \sim E/M_h$.}
\label{fig:R1}
\end{figure*}

\bigskip

To illustrate the emergence of Higgsplosion, we plot ${\cal R}_n(E)$ of \eqref{eq:Rnp2} 
in Fig.~\ref{fig:R1} at fixed values of $E$ and vary $n$. 
The values of the energy are chosen to
zoom on the range where ${\cal R}_n(E)$ changes from the exponentially small to exponentially large values.
The energy dependence of this transition is very sharp, this fact playing an important role in effectively cutting off at $E_c$
the loop integrals contributing to the Higgs mass in the solution to the Hierarchy problem proposed in \cite{Higgsplosion}.
It is also easy to understand the peak in the particle number $n$ for each fixed-energy plot in Fig.~\ref{fig:R1}.
At relatively low values of $n$ the multi-particle rate is small, as expected, while at the maximal value 
$n_{\rm max}=E/M_h$ the rate is zero again as we have run out of the phase space for the final-state particles;
hence the local maximum in $n$ appears before the edge of the phase-space is reached, and is located 
at the values of $n$ parametrically close to the maximal $n$.

\bigskip

The expression for the multi-particle rate \eqref{eq:Rnp2} should of course not be taken as the
full final result for the physical Higgsplosion rate. We have already emphasised that this result is an
approximation derived in the simplified scalar model \eqref{eq:L} and in the simplifying non-relativistic limit.
Specifically, our main result \eqref{eq:gln} was derived on the multi-particle threshold, i.e. at $\varepsilon=0$.
Hence the higher-order corrections in $\varepsilon$ will be present in the expression for the rate 
in the $\lambda n$ limit. Denote these corrections $f_{\lambda n; \varepsilon} ({\lambda n, \varepsilon} )$, so that
\[ \Delta_{\rm new} W \,=\, \frac{\lambda n}{\lambda}\,\, f_{\lambda n; \varepsilon} ({\lambda n, \varepsilon} )\,,
\]
and the now modified rate becomes,
\[
{\cal R}_n(E)\,\sim \,  \int_0^{\varepsilon_{nr}} d \varepsilon\, \left(\frac{\varepsilon}{3 \pi}\right)^{\frac{3n}{2}} 
\exp \left[ n\, \left( 
0.85\, \sqrt{\lambda n} \,+\, \log \lambda n\,+\,f_{\lambda n; \varepsilon} ({\lambda n, \varepsilon} )
\,+\, c
\right)\right] 
\label{eq:RnpN}
\]
where we have included the new correction $ n \sim f_{\lambda n; \varepsilon} ({\lambda n, \varepsilon} )$ 
and have also made explicit the fact that the $3n/2 \log \varepsilon/(3 \pi)$ factor in
the exponent of the rate \eqref{eq:Rnp2} originated from the integration 
over the non-relativistic $n$-particle phase-space with a cut-off at $\varepsilon_{nr} < 1$.
The constant $c$ absorbs various constant factors appearing in the original rate.

The integral above is of course meant to be computed in the large-$n$ limit by finding the
saddle-point value $\varepsilon=\varepsilon_{\star}$. The main point of the exercise is to determine
(1) whether there is a regime where $\varepsilon_{\star} \ll 1$ so that our near-the-threshold approach is justified,
and (2) whether the saddle-point value of the rate itself is large. These requirements should 
tell us something about the function $f_{\lambda n; \varepsilon}$.

Let us assume that the correction to our result has the form,
\[
f_{\lambda n; \varepsilon} ({\lambda n, \varepsilon} ) \,=\, - a\, \varepsilon \, (\lambda n)^p
\,, \label{eq:fnew}
\]
where $a$ and $p$ are constants. This function is supposed to represent the higher-order in 
$\varepsilon$ correction to our result in the small-$\varepsilon$, large-$\lambda n$ limit.
The integral we have to compute is,
\[
{\cal R}_n \,\sim \,
e^{n\, \left( 0.85\, \sqrt{\lambda n} \,+\, \log \lambda n \,+\, \tilde c
\right)}\,\,
\int d \varepsilon\,
e^{n\, \left( \frac{3}{2} \log \varepsilon \,-\, a\, \varepsilon \, (\lambda n)^p\right) } \,.
\label{eq:RnpNsp}
\]
Denoting the $\varepsilon$-dependent function in the exponent $s(\varepsilon)$,
\[ s(\varepsilon) \,=\, \frac{3}{2} \log \varepsilon \,-\, a\, \varepsilon \, (\lambda n)^p\,,
\]
we can compute the saddle-point,
\[ \frac{\partial s(\varepsilon)}{\partial \varepsilon} \,=\, 0  \quad \Rightarrow \quad
\varepsilon_{\star} \,=\, \frac{3}{2}\, \frac{1}{a} \, \frac{1}{(\lambda n)^p}\,,
\]
and the value of the function $s$ at the saddle-point,
\[
s(\varepsilon_{\star}) \,=\, -\frac{3}{2}\, \left(p \log \lambda n \,+\,1\,-\, \log \frac{3}{2a} \right)\,.
\]
Combining this with the function in the exponent in front of the integral in \eqref{eq:RnpNsp}
we find the saddle-point value of the rate,
\[
{\cal R}_n (\varepsilon_\star) \,\sim \,
\exp \left[n\, \left( 0.85\, \sqrt{\lambda n} \,-\, \left(\frac{3p}{2}-1\right) \log \lambda n \,+\, {\rm const} \right)
\right]
 \,.
\label{eq:RnpNsp2}
\]
This is the value of the rate at the local maximum, and since the factor of $\sqrt{\lambda n}$ grows faster
than the $-\log \lambda n $ term, the peak value of the rate is exponentially large
in the limit of $\sqrt{\lambda n} \infty$. It is also easy to verify that this conclusion is
consistent within the validity of the non-relativistic limit.
In fact, the value of $\varepsilon$ at the saddle-point is non-relativistic,
\[
\varepsilon_{\star} \,=\, \frac{3}{2}\, \frac{1}{a} \, \frac{1}{(\lambda n)^p} \, \to\, 0 \,,
\quad {\rm as} \quad 
\lambda n \to\, \infty
\,.
\]
We thus conclude that the appearance of the higher-order in $\varepsilon$ corrections to our result 
in the form \eqref{eq:fnew} do not prevent the eventual Higgsplosion in this model at  least in the
formal limit $\sqrt{\lambda n} \to \infty$ where we have found that
\[
{\cal R}_n (\varepsilon_\star) \, \gg \, 1\,.
\]
The growth persists for any constant values of $a$ and $p$. In fact, if $a$ was negative, the
growth would only be enhanced. In   \eqref{eq:fnew} we have assumed that the function 
goes as $\varepsilon$ to the first power. The higher powers would not change the conclusion, while
the effect of $\sim \varepsilon^0$ is what is already taken into account in \eqref{eq:gln}.

\bigskip

For completeness, we note that only a rather extreme type of corrections  would prevent the Higgsplosion in this theory.
They would have to be of the form, 
\[
f_{\lambda n; \varepsilon} ({\lambda n, \varepsilon} ) \,=\, - \, \varepsilon \, e^{(\lambda n)^p}
\,, 
\]
which in terms of ${\cal R}_n$ would amount to a negative double exponential,
\[ {\cal R}_n \,\sim\, \exp\left[-n \varepsilon \, e^{(\lambda n)^p} \right]
\,\sim\, \exp\left[-E \, e^{(\lambda n)^p} \right]
\,,
\]
which we find to be rather unlikely.

\bigskip

Our discussion up to now concentrated entirely on a simple scalar field model.
If more degrees of freedom were included, for example the $W$ and $Z$ vector bosons and
the SM fermions, new coupling parameters (such as the gauge coupling $\alpha_w$ and the top Yukawa $y_t$) 
would appear in the expression for the rate along with the final state particle multiplicities. As there are more parameters,
the simple scaling properties of ${\cal R}_n$ in the pure scalar theory will be modified. If the scaling persists, there
will be more to it than the two variables $\lambda n$ and $\varepsilon$. Understanding of how this works and 
investigating the appropriate weak-coupling / high-multiplicity semiclassical limit or limits is an important task for 
future work.

One can however consider such effects in the leading order in the loop expansion, i.e. where the  $\lambda n $
parameter is considered to be small. Very recently the contributions of virtual top quarks (more generally, fermions and/or
scalars coupled to the Higgs) to the multi-Higgs amplitudes on the threshold  were computed in
\cite{Voloshin:2017flq}. Their result for the case of the top quark is that the threshold amplitude of the pure Higgs theory
is multiplied by an overall factor,
\[
{\cal A}_n\, \longrightarrow\,  {\cal A}_n\,\left(1\,-\, C\, n^{2} \lambda \, 
\frac{1}{n^{6-4\frac{m_t}{M_h}}} \,+\, {\cal O}(\lambda^2) \right)\,,
\label{eq:Vmt}
\]
where $m_t$ is the top quark mass and the numerical coefficient $C$ for the top-quark correction is
\[
C\,=\, C\left(\frac{m_t}{M_h}\right)\, \simeq\,
(8.0 + i\, 5.8) \frac{\sqrt{3}}{8 \pi}\,.
\]
What is currently unknown is whether these corrections can exponentiate, and 
if so, what their effect might be in the appropriate large $\lambda n$ limit.
If there is no effective exponentiation of these effects, in our view it would be extremely unlikely to expect that a
precise cancellation in the prefactor of the multi-particle rate could occur. If the exponentiation 
of these effects does occur, as it did for the virtual corrections within the scalar sector itself,  
the possible effects of it need yet to be understood.
We have written the top-quark correction in \eqref{eq:Vmt}  in a suggestive form, singling out the factor of $n^2 \lambda$.
This was done in order to compare its effect to the leading order correction to the ${\cal R}_n$
in the scalar theory ({\it cf} \eqref{eqnl}),
\[
n^2 \lambda\,  \frac{\sqrt{3}}{4 \pi}\, \qquad vs. \qquad 
 n^{2} \lambda \, 
\frac{1}{n^{6-4\frac{m_t}{M_h}}} \, 2{\rm Re}C\,\simeq\, \frac {n^{2} \lambda}{n^{0.48}}\, \frac{8\sqrt{3}}{4 \pi}
\]
Formally, there is a parametric suppression of the top-corrections relative to the leading order loop correction
in the scalar sector. In the asymptotic limit of large $n$ they would be subleading. The resummation of the
loop corrections in the scalar sector is what have resulted in the large $\sqrt{\lambda n}$ effect we have computed.
On the other hand, at present little is known about the prospects of resummation or even the sign of the effect 
related to the top quark corrections. Similarly, important effects should also from including the vector bosons, 
and these avenues should be pursued in future. 

\bigskip
\bigskip

\noindent The main conclusion we draw from the results presented in this paper is that we have demonstrated 
that the Higgsplosion phenomenon is realised
above a critical energy/high-multiplicity scale in a concrete QFT settings.
The theory we used is the scalar QFT  \eqref{eq:L} with the spontaneous symmetry breaking. 
The idea of Higgsplosion as a possible solution to the Higgs mass-induced Hierarchy problem also has direct model-building and phenomenological consequences. 
%
From the phenomenological perspective the idea is also testable, for example by studying the feasibility of
observing of the muti-Higgs and multi-vector-boson production cross-sections at future hadron colliders
\cite{VVK2,Degrande:2016oan}. The Higgsplosion yield at colliders
was recently addressed in Ref.~\cite{Gainer:2017jkp}.
We believe that studies of Higgsplosion phenomenology offers promising and exciting opportunities for future
in particle physics and cosmology.

\bigskip

\section*{Acknowledgements}

I am grateful to Joerg Jaeckel and Michael Spannowsky for many useful discussions.
This work is supported by the STFC through the IPPP grant and  
by the European UnionÕs Horizon 2020 research and innovation programme under the Marie 
Sklodowska-Curie grant agreement No 690575.

\bigskip

\bibliographystyle{h-physrev5}

\end{document}